\documentclass[aps,prl,twocolumn,preprintnumbers,groupedaddress]{revtex4}

\usepackage{amssymb}
\usepackage{epsfig} 
\usepackage{amsmath,color} 
\usepackage{graphicx} 
\usepackage{ulem}

\usepackage{wrapfig}

\newcommand{\beq}{\begin{equation}}
\newcommand{\eeq}{\end{equation}}
\newcommand{\bea}{\begin{eqnarray}}
\newcommand{\eea}{\end{eqnarray}}
\newcommand{\ba}{\begin{array}}
\newcommand{\ea}{\end{array}}
\newcommand{\bi}{\begin{itemize}}
\newcommand{\ei}{\end{itemize}}
\newcommand{\bn}{\begin{enumerate}}
\newcommand{\en}{\end{enumerate}}
\newcommand{\bc}{\begin{center}}
\newcommand{\ec}{\end{center}}
\renewcommand{\l}{\left}
\renewcommand{\r}{\right}

\newcommand{\eq}[1]{Eq.~(\ref{#1})}

\newcommand{\GeV}{\mathinner{\mathrm{GeV}}}

\begin{document}


\title{A double-running inflaton mass for a flat potential and an assisted hilltop inflation}


\author{Wan-Il Park}
\email[]{wipark@jbnu.ac.kr}
\affiliation{
Division of Science Education and Institute of Fusion Science, Chonbuk (Jeonbuk) National University, Jeonju 561-756, Korea}


\date{\today}

\begin{abstract}
We propose a scenario to flatten inflaton potential for a small-field slow-roll inflation from the origin.
In the scenario, inflaton mass-square gets through a renormalization group running which depends on not only inflaton field but also an assisting field.
Thanks to the assisting field participating in the running, initial condition for a slow-roll inflation can be set naturally, and inflaton potential can be flattened by the vacuum expectation value of the field.
Applying this idea, we propose a scenario of inflation, dubbed here as \textit{assisted hilltop inflation}, which is free from the initial condition and the flatness problems of inflation.
\end{abstract}


\maketitle


\section{Introduction}

As a modern paradigm of cosmology, inflation \cite{Guth:1980zm,Sato:1980yn,Starobinsky:1980te} provides a very simple solution to the problems of hot Big-Bang cosmology, while providing a compelling way to generate seeds of the large scale structure in the present universe \cite{Mukhanov:1981xt,Mukhanov:1982nu}.
Such an attractive property is based on so-called slow-roll approximation which requires a flat inflaton potential having a curvature smaller than the expansion rate by about one order of magnitude during inflation.
It's been, however, recognized that the slow-roll approximation is in practice quite non-trivial to be realized in high energy theories, as on general grounds the inflaton receives Hubble scale mass contribution(s) caused by Planck-suppressed operator(s).
In particular, in the framework of supergravity which is believed to be the low energy effective theory of string theories and the proper UV framework for inflation, the problem  manifests itself in the form of supergravity ($F$-term) scalar potential \cite{Copeland:1994vg,Stewart:1994ts}.
The slow-roll approximation (especially the smallness of the inflaton mass-square relative to the expansion rate during inflation) is clearly broken in this case and a nearly scale-invariant CMB power spectrum can not be explained.
This is the well-known flatness problem of the inflaton potential, also known as the `$\eta$-problem' from the second slow-roll parameter.  

There have been a variety of ideas to resolve or avoid the $\eta$-problem in constructions of inflation models (see, for example, Ref.~\cite{Yamaguchi:2011kg}).
In regard to $F$-term inflation subject to the $\eta$-problem, most of them use special patterns of K$\ddot{a}$hler potential and superpotential under an assumption of a certain symmetry.
Introducing a new symmetry (for example, shift symmetry \cite{}) may be a convenient way to remove dangerous gravitational corrections to the inflaton mass.
But it is not clear if there is any symmetry other than gauge symmetry, which is free from gravitational breaking.
Hence, such an introduction of a symmetry may not be a solution to the problem.

It may be thought that $\eta$-problem can be easily solved if one allows trans-Planckian excursions of inflaton, e.g., like in \textit{natural inflation} \cite{Freese:1990rb}.
However,  such a large excursion of inflaton is generically problematic in view of the validity of a given theory at hand, since higher order terms suppressed by Planck-scale are out of control.
A possible way of circumventing this issue may be from the fact that generically inflaton lives in a multi-dimensional field space.
In other words, inflaton doesn't need to be a fixed (straight) coordinate but can be just a non-trivial trajectory in the field space. 
In principle, the trajectory can be compactified in a sub-Planckian region even though its length is much larger than Planck scale \cite{Silverstein:2008sg,McAllister:2008hb,Berg:2009tg,McDonald:2014oza,McDonald:2014nqa,Li:2014vpa,Carone:2014cta,Barenboim:2014vea,Barenboim:2015zka,Barenboim:2015lla} (see also \cite{Kim:2004rp}).
Notably, the resulting inflation can be effectively the same as a single-field slow-roll inflation.
The elongation of the trajectory does again require a tuning of parameters, so it may not look improving the original problem in a natural manner, but at least it provides a way of obtaining large-field inflation scenarios which are free from the trans-Planckian issue.

Considering a non-minimal coupling to gravity is another approach \cite{Bezrukov:2007ep} although its original motivation is not on the problem.
Depending on setups, the Jordan frame potential can be flattened or amplified in Einstein frame, and $\eta$-problem becomes irrelevant.
However, even if the question on the origin of the \textit{large} non-minimal coupling is put aside, there seems still to be issues of fine-tuning higher order terms and unitarity \cite{Burgess:2010zq} (see, however, \cite{Giudice:2010ka} for example).

In the conventional Einstein gravity, when it is difficult to block dangerous gravitational corrections to the inflation mass in the framework of SUGRA, it might be natural to consider the smallness of $\eta$ as a result of some dynamics.  
In this regard, an interesting and seemingly natural approach to $\eta$-problem is to use the renormalization group running of the inflaton mass, as discussed in Ref.~\cite{Stewart:1996ey}.
The idea was accommodated in Hybrid inflation to match observations \cite{Stewart:1997wg}, but a recent analysis showed that several well-known models suffer from either blue spectrum or too large density perturbation caused by entropy perturbation \cite{Clesse:2013jra}.
Instead of inward inflaton, it could be possible to use outward direction for inflation, but such a case suffers from the issue of initial condition and too much $e$-foldings.

In this work, inspired by the idea of \textit{running mass inflation (RMI)}, we propose a new way to flatten inflaton potential, in which two flat-directions participate in the renormalization group running of the inflaton mass-square.
The novel aspect of this idea is that, without resorting to any ad hoc symmetry in K$\ddot{\rm a}$hler potential, the mass-square ($m_\phi^2$) of inflaton (say $\phi$, one of the flat-directions) can be dynamically adjusted to be small by a vacuum expectation value (VEV) of the other flat direction (say $\chi$), and the running of $m_\phi^2$ along $\phi$ appears only after crossing the cutoff scale set by the VEV of $\chi$. 
This idea allows a natural realization of a hilltop inflation \cite{Boubekeur:2005zm} along $\phi$, which we may call as \textit{assisted hilltop inflation}.

This work is organized as follows.
We first introduce the original idea of RMI briefly and then describe the RG-running of the inflaton mass along two flat directions.
Subsequently, we will consider a concrete model of inflation, and show a result of a numerical analysis as an example.
Finally, the conclusion will be drawn.

\section{Field-dependent running mass}
In the framework of supergravity (SUGRA), the defining energy scale of a theory or model is assumed to be the reduced Planck scale ($M_{\rm P} \simeq 2.4 \times 10^{18} {\rm GeV}$), the fundamental scale of 4-D SUGRA.
Now, suppose we are interested in an inflation with a quadratic potential of a scalar field $\phi$ representing the canonically normalized radial component of a flat direction, 
\beq \label{eq:V}
V = V_0 + \frac{1}{2} m^2 \phi^2 + \dots
\eeq
where $m^2$ is the soft SUSY-breaking mass-square parameter of $\phi$, and $\dots$ represents possible higher order terms including terms stabilizing $\phi$ in case of $m^2<0$ all the way to Planck scale.
$V_0$ is set for vanishing cosmological constant at zero-temperature true vacuum.
At the defining scale, the general form of K$\ddot{a}$hler potential with $\mathcal{O}(1)$ numerical coefficients leads to \cite{Copeland:1994vg,Stewart:1994ts}
\beq
m^2 = \mathcal{O}(3 H^2)
\eeq
where $H$ is the expansion rate of the universe when $V_0$ dominates the universe.
Hence the second slow-roll parameter, $\eta \equiv M_{\rm P}^2 V''/V$ with $V''$ the second derivative with respect to $\phi$, appears to be of order unity with either sign allowed.
This is the heart of $\eta$-problem, and a slow-roll inflation is not realized.
However, if $\phi$ has interactions with light (relative to the energy scale of interest) particles, the mass-square parameter will get through renormalization group (RG) running so that $m^2$ depends on the renormalization scale ($Q$) or the energy scale of interest. 
Now, suppose particles coupled to $\phi$ obtain masses proportional to $\phi$ which is much larger than the scale of soft SUSY-breaking mass parameters (say $m_{\rm soft}$).
Then, assuming the strength of the couplings to be of order unity for simplicity, the renormalization will be cutoff at the scale determined by $\phi$.
Hence, setting $Q=\phi$, one finds field-dependent mass-square $m^2(\phi)$.
From the renormalization group equations of $m^2$ and the mass parameters of coupled fields, in principle one can find the form of $m^2(\phi)$.
Here, for simplicity, we consider 1-loop RG-running of $m^2$ and ignore the scale dependence of the beta-function of the RG-equation.
Then, one can write $m^2(\phi)$ as
\beq \label{1loop-mphisq}
m^2(\phi) = m_0^2 + \beta_m \ln \l( \frac{\phi}{M_{\rm P}} \r)
\eeq 
where $\beta_m \equiv d m^2/ d \ln Q$, the beta-function of $m^2$.
We assume both $m_0^2$ and $\beta_m$ negative in the following discussion. 
In the vicinity of a field value $\phi_\wedge (\gg m_{\rm soft})$, one can rewrite $m^2(\phi)$ as
\beq
m^2(\phi) = m_\wedge^2 + \beta_m \ln \frac{\phi}{\phi_\wedge}
\eeq
where $m_\wedge^2 \equiv m^2(\phi_\wedge)$.
In this approximation, ignoring the running of $\beta_m$ and higher order terms of $V$ in \eq{eq:V}, one finds derivatives of $V$ as 
\bea \label{dV}
V' &=& \l[ m_\wedge^2 + \beta_m \l( \frac{1}{2} + \ln \frac{\phi}{\phi_\wedge} \r) \r] \phi 
\\ \label{ddV}
V'' &=& \beta_m + \l[ m_\wedge^2 + \beta_m \l( \frac{1}{2} + \ln \frac{\phi}{\phi_\wedge} \r) \r]
\eea
Thus, if $V'(\phi_\wedge) = 0$, 
\beq
m_\wedge^2 = - \frac{1}{2} \beta_m
\eeq
and
\beq
V''(\phi) = \beta_m \l( 1 + \ln \frac{\phi}{\phi_\wedge} \r)
\eeq
Typically, $\beta_m /m_0^2= \mathcal{O}(10^{-2})$.
Hence,  if the field value $\phi_*$ associated with the Planck pivot scale is close to $\phi_\wedge$, i.e., $\phi_* \sim \phi_\wedge$, a flat enough power spectrum matching Planck data can be achieved in a natural way.
Inflation can take place along either side of $\phi_\wedge$, but a recent analysis showed that the case of $\beta_m < 0$ with inflation for $\phi_\wedge < \phi$ fit better to Planck data \cite{Martin:2013tda}. 
There are however two issues in this case: (i) the initial condition, i.e., setting $\phi \sim \phi_\wedge$ from the beginning, and (ii) too much $e$-foldings to the end of inflation.
The former may require eternal inflation \cite{Vilenkin:1983xq,Guth:1985ya,Linde:1986fc,Linde:1986fd,Starobinsky:1986fx} or tunneling to populate $\phi$ around the local maximum of the potential.
The latter requires an additional feature to reduce $e$-foldings.

\section{Double-running of mass parameter}

The RG-running of the inflaton mass-square parameter depends on masses of particles having a sizable coupling to the inflaton.
In a simple case, their masses depend on the field value of inflaton as used in RMI scenario.
However, those particles may also interact with other fields which develop large non-zero VEVs, obtaining additional masses.
If such a mass-generation takes place before the last inflation along $\phi$ takes place, there will be threshold effects or cutoff in the RG-running of inflaton mass-square along $\phi$.

As an example, we consider two orthogonal flat directions $\chi$ and $\phi$ consisting of gauge-charged fundamental fields and assume that those field consisting of $\chi$ share some gauge-interactions with the fields consisting of $\phi$.  
At the defining scale, for a generic form of K$\ddot{a}$hler potential in SUGRA $\chi$ and $\phi$ are expected to have a Hubble scale soft SUSY-breaking mass-square with the possibility of either sign.
Those mass parameters are subject to renormalization-group (RG) runnings in the presence of gauge and Yukawa couplings if coupled particles are light relative to the energy scale of interest.
We assume that $\chi$ has a negative mass-square parameter for the whole range of the field and is stabilized by a higher order term.
On the other hand, for $\phi$ we assume that the mass-square parameter is negative at the input scale (i.e., Planck scale) and subject to a significant negative RG-running as the case of RMI scenario.
Then, ignoring the possible phase-dependent term(s) of $\chi$ and $\phi$ for simplicity, and using the same notation for their canonically normalized scalar radial components, the potential involving them during inflation can be written as
\beq
V = V_0 + V_\chi(\chi) + V_\phi(\chi,\phi)
\eeq
with
\bea \label{eq:Vchi}
V_\chi &=& \frac{1}{2} m_\chi^2 \chi^2 + \l( \frac{\lambda_\chi \chi^{q-1}}{M_{\rm P}^{q-3}} \r)^2
\\ \label{eq:Vphi}
V_\phi &=& \frac{1}{2} m_\phi^2 \phi^2 + \dots 
\eea
where $\lambda_\chi$ is a dimensionless coefficient of $\mathcal{O}(1)$ at most, and $q \geq 4$ as an integer.
The mass parameters are taken to be
\bea
m_\chi &=& c_\chi m_0^2
\\ \label{mphi-run}
m_\phi^2 &=& - c_\phi m_0^2 + \frac{1}{2}\beta_m \ln \l( \frac{\chi^2 + \phi^2}{m_{\rm soft}^2} \r)
\eea
where $c_\chi \sim c_\phi \sim \mathcal{O}(1) > 0$ are numerical coefficients, $m_0^2(<0)$ represents a negative Hubble scale mass-square parameter defined at Planck scale, $\beta_m \equiv d m_\phi^2 / d \ln Q$ is determined by loop-corrections in a given model, and $m_{\rm soft}$ is a typical mass scale induced by the supersymmetry-breaking effect of the energy density for inflation.
Even if it may start with negative value, $m_\phi^2$ can run to a positive value (which we set as $-c_\phi m_0^2$) around the origin due to a rather strong running.
We assume this is the case.
Note that $m_\phi^2$ depends on not only $\phi$ but also $\chi$.
This is possible when the RG-running is dominated by gauge interactions shared by both $\chi$ and $\phi$.

A plausible thermal history in our scenario is as follows.
At high temperature, the interactions to thermal bath can hold $\chi$ and $\phi$ around the origin by providing them large thermal masses.
As time goes on, due to its negative mass-square around the origin, $\chi$ rolls out.
This makes particles (for example, gauge/gaugino fields interacting with $\chi$) obtain $\chi$-dependent masses and causes RG-running of $m_\phi^2$ along $\chi$. 
As a result, if $\chi_0$ (the VEV of $\chi$) is far away from the origin, it is possible that the local minimum around the origin along $\phi$ turns to a local maximum as $\chi$ acrosses $\chi_\times$ satisfying
\beq \label{chiX}
\frac{\beta_m}{c_\phi m_0^2} \ln \l( \frac{\chi_\times}{m_{\rm soft}} \r) = 1
\eeq
in 1-loop approximation.
Note that, although it is not automatic, $\chi_0$ can be determined by the interplay of the quadratic term and a higher order term so as to be $\chi_0 \sim \chi_\times$ by a reasonable choice of parameters.
Then, in the vicinity of $\chi_\times$ one can rewrite $m_\phi^2$ as
\beq
m_\phi^2 = \frac{1}{2} \beta_m \ln \l( \frac{\chi_0^2 + \phi^2}{\chi_\times^2} \r)
\eeq
For $\chi_0 > \chi_\times$, as $\chi$ across $\chi_\times$, $\phi$ starts rolling out slowly.
For $\phi \ll \chi_0$, the curvature along $\phi$ is found to be
\beq \label{ddVphi}
\frac{d^2 V_\phi}{d \phi^2} \simeq \beta_m \l[ \ln \frac{\chi_0}{\chi_\times} + 3 \l( \frac{\phi^2}{\chi_0^2} \r) \r]
\eeq
i.e., the curvature along $\phi$ is dominantly controlled by the ratio, $\chi_0/\chi_\times$.
Hence, in addition to a non-slow-roll inflation along $\chi$, it is possible to realize a slow-roll inflation along $\phi$ with the spectral index under control.
Note that in this type of inflation the initial condition for inflation can be set naturally by an earlier stage of thermal inflation realized at the symmetry-enhanced point (i.e., the origin in the current scenario) in the field space. 

In regard to $\eta$-problem, the smallness of $\eta$ is now from the closeness of $\chi_0$ to $\chi_\times$.
It may be still regarded as a tuning, but $\chi_0$ can fall in the vicinity of $\chi_\times$ for a reasonable choice of parameters, making inflaton mass-square small even if it is of Hubble scale at the defining scale of the theory.

\section{A model}

\begin{figure*}[ht]
\begin{center}
\includegraphics[width=0.49\linewidth]{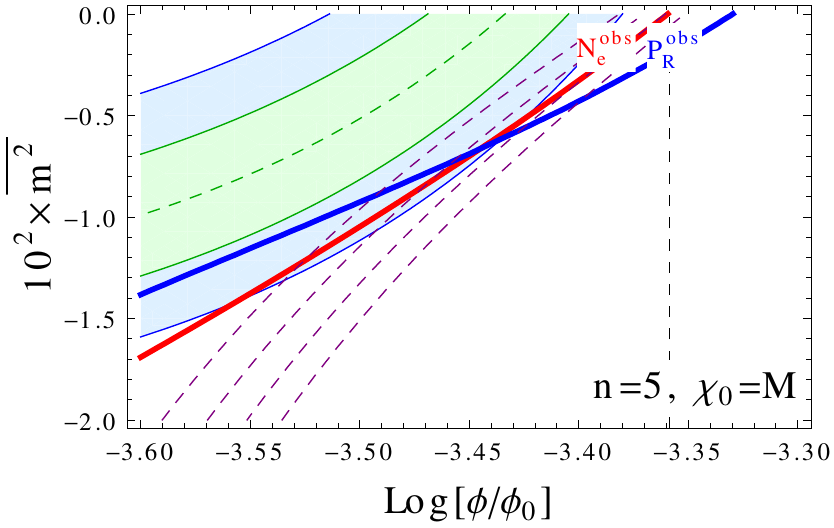}
\includegraphics[width=0.49\linewidth]{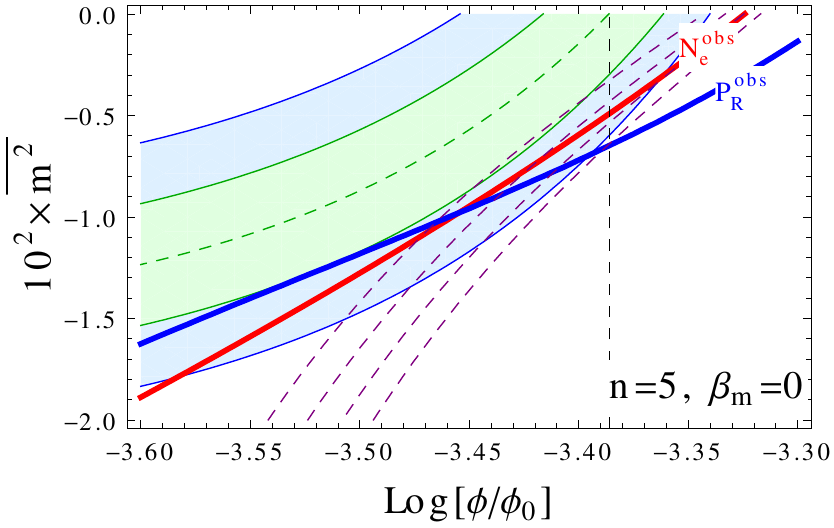}
\caption{Parameter space matching observations for $n=5, M=2.3 \times 10^{12} \GeV$, and $\phi_0 = 10^{15} \GeV$.
\textit{Left}: $\chi_0=M$ with $\beta_m = 1/\ln(\chi_0/M_{\rm P})$,
\textit{Right}: $\beta_m=0$.
In both panels, green and blue bands are 1- and 2-$\sigma$ allowance of spectral index $n_s$, blue line is for $P_R = P_R^{\rm obs}=2.1 \times 10^{-9}$ \cite{Ade:2015lrj}, red line is for $N_e = N_e^{\rm obs}$, and purple dashed lines are for $10^4 \times d n_s / d \ln k = 7,8,9,10$ from left to right. 
The vertical dashed line stands for $\phi_*$ in the case of a pure hilltop potential without quadratic term. 
For the estimation of $\Gamma_I$ in \eq{eq:Gamma-I} we took $\gamma_I=200$.
}
\label{fig:para-space-n5}
\end{center}
\end{figure*}
In the mechanism discussed in the previous section, for the last inflation along $\phi$ the role of $\chi$ can be simply replaced by a cutoff scale in the RG-running of the inflation mass-square.
Keeping this in mind, we consider an inflaton potential along $\phi$,
\beq \label{eq:V}
V(\phi) = \frac{1}{2} m_\phi^2 (\phi) \phi^2 + M^4 \l[ 1 - \l( \frac{\phi}{\phi_0} \r)^n \r]^2
\eeq
where 
\beq \label{eq:m-phi}
m_\phi^2 (\phi) = m_c^2 + \frac{1}{2} \beta_m \ln \l( 1 + \frac{\phi^2}{\chi_0^2} \r)
\eeq
with $m_c^2 \equiv \beta_m \ln (\chi_0/\chi_\times)$, and $n (\geq 4)$ is an integer.
The scale of $M$ and $\phi_0$ are treated as free parameters here, but satisfying $\chi_0 \ll \phi_0 \ll M_{\rm P}$.
For $\phi_*$ associated with a cosmological scale in observations, if $\phi_* \lll \chi_0$, current scenario would produce essentially the same inflationary observables as the well-studied model with the quadratic term replaced by a tree-level mass term \cite{Kohri:2007gq}.
However, $\chi_0 (\sim \chi_\times)$ is not arbitrarily large for a reasonable choice of $\beta_m$.
Actually, it may be reasonable to expect for it to be of an intermediate scale.
In this case, the effect of the RG-running of $m_\phi$ along $\phi$ may have a sizable impact on inflationary observables.
When the $\beta_m$ term in \eq{eq:m-phi} is non-negligible, even if slow-roll parameters and inflationary observables from the potential in \eq{eq:V} are easily found, an analytic expression of $e$-foldings is non-trivial to be obtained.
Also, note that the amount of $e$-foldings required to match observations is constrained by the energy density of inflation and the reheating temperature \cite{Lyth:1998xn}.
In order to take this into account, we take the decay rate of inflaton for $M \ll \phi_0$ to be
\beq \label{eq:Gamma-I}
\Gamma_I = \frac{\gamma_I}{8 \pi} \frac{m_I^3}{\phi_0^2}
\eeq
where $\gamma_I$ is a numerical constant taking into account of decay channels and couplings to light particles, and $m_I \approx \sqrt{2} n M^2/\phi_0$.

In Fig.~\ref{fig:para-space-n5}, as an example, we depicted a parameter space matching observations for $n=5$ with and without $\beta_m$ contribution in \eq{eq:m-phi}.
Comparing two panels in the figure, one can see how large the impact of $\beta_m$ on the spectral index can be.
It depends on how large or small $\chi_0/M$ is though.
As expected from the parametric dependence of power spectrum and $e$-foldings on $M$ and $\phi_0$, it is found that for a larger $M$, while the change of the red line is minor, the blue line is pushed downward significantly, going out of the 2-$\sigma$ allowed band.
Such a pattern can be compensated by taking a larger $\phi_0$, but then the redline is pushed down significantly.
Hence, it turned out that there are no large rooms to adjust $M$ and $\phi_0$ (which are inter-related) for a given set of the other parameters. 
On the other hand, taking a large $\chi_0$ pushes both red and blue lines toward the best-fit region.
This is because, as $\chi_0$ becomes larger, the impact of mass-running in the vicinity of $\phi_*$ becomes weaker.
For a larger $n$, e.g., $n=6$, the allowed parameter space can fall into to the 1-$\sigma$ best-fit region of the spectral index.

It should be noted that in a concrete realization of a hilltop potential $\phi_0$ may not be an arbitrarily free parameter.
For example, the potential we are considering here might be from a superpotential
\beq \label{eq:W}
W = S \l( M^2 - \frac{\lambda_\phi \phi^n}{n M_*^{n-2}} \r)
\equiv M^2 S \l[ 1 - \l( \frac{\phi}{\phi_0} \r)^n \r] 
\eeq
where $S$ is a gauge singlet superfield, $\phi$ is assumed to be a hidden sector flat-direction something like a MSSM $D$-flat-direction, $\lambda_\phi$ is a numerical coupling constant of order unity or less, $M_*$ is a UV-cutoff scale which may be either GUT or Planck scale, and $\phi_0^n \equiv n M^2 M_*^{n-2}/ \lambda_\phi$.
Hence, $\phi_0$ depends dominantly on $M_*$ and $M$ as the next for $n \geq 5$.
Such a form of superpotential may be from, for example, a proper choice of gauged $U(1)_R$ symmetry \cite{Cremmer:1983yr,Chamseddine:1995gb}.
From numerical tests we found that for $\chi_0 \sim \mathcal{O}(10^{12-14}) \GeV$ it is possible to find a set of working parameters for $M_* = M_{\rm GUT}$, but difficult for $M_* = M_{\rm P}$.

\section{Conclusions}

In this work, we proposed a simple mechanism to flatten inflaton potential when gravitational corrections to K$\ddot{a}$hler potential add Hubble scale mass to inflaton.
This mechanism is based on the renormalization group running of the inflaton mass-square but the running depends on not only inflaton ($\phi$) but also an assisting field ($\chi$) which are flat-directions made of several gauge-charged fundamental fields.
The curvature of the potential along inflaton direction can be adjusted to be small dominantly by the vacuum expectation value of the assisting field.

As a concrete realization of inflation utilizing this mechanism of double-running inflaton mass, we considered a hilltop inflation scenario, dubbed as \textit{assisted hilltop inflation}, where the inflaton potential consists of a quadratic term, whose mass parameter is subject to RG-running, and a typical hilltop potential in addition to a potential for the assisting field only.
Contrary to known hilltop inflation scenarios without the running effect of mass parameter taken into account, assisted hilltop inflation does not suffer from the initial condition and flatness problems.
It requires just an initial hot universe.    
Assisted hilltop inflation may be regarded as a UV-completion of some of hilltop inflation scenarios.

The idea of a double-running inflaton mass presented here does not automatically set inflaton mass small relative to the expansion rate during inflation, but it provides a simple way of obtaining a small curvature of inflaton potential for a reasonable choice of parameters.

\section{Acknowledgements}
The author thanks Gabriela Barenboim and Jinsu Kim for fruitful discussions at the early stage of this work.
This work is supported by research funds for newly appointed professors of Chonbuk National University in 2017, and by Basic Science Research Program through the National Research Foundation of Korea (NRF) funded by the Ministry of Education (No. 2017R1D1A1B06035959).

\end{document}